\documentclass[fleqn,twoside]{article}
\usepackage{espcrc2}
\usepackage{epsfig}

\usepackage{graphicx}
\usepackage[figuresright]{rotating}

\newcommand{\AmS}{{\protect\the\textfont2
  A\kern-.1667em\lower.5ex\hbox{M}\kern-.125emS}}

\title{The Laser Astrometric Test of Relativity Mission}

\author{Slava G. Turyshev\address[JPL]{Jet Propulsion Laboratory, California Institute
of  Technology, \\4800 Oak Grove Drive, Pasadena, CA 91109},
        Michael Shao\addressmark ~and
        Kenneth L. Nordtvedt, Jr.\address{Northwest Analysis, 118 Sourdough Ridge Road, Bozeman MT 59715 USA}
        }
       
\begin{document}

\begin{abstract}
This paper discusses new fundamental physics experiment to test relativistic gravity at the accuracy better than the effects of the 2nd order in the gravitational field strength, $\propto G^2$. The Laser Astrometric Test Of Relativity (LATOR) mission uses laser interferometry between two micro-spacecraft whose lines of sight pass close by the Sun to accurately measure deflection of light in the solar gravity.  The key element of the experimental design is a redundant geometry optical truss provided by a long-baseline (100 m) multi-channel stellar optical interferometer placed on the International Space Station (ISS). The interferometer is used for measuring the angles between the two spacecraft.
	In Euclidean geometry, determination of a triangle's three sides determines any angle therein; with gravity changing the optical lengths of sides passing close by the Sun and deflecting the light, the Euclidean relationships are overthrown. The geometric redundancy enables LATOR to measure the departure from Euclidean geometry caused by the solar gravity field to a very high accuracy.
	LATOR will not only improve the value of the parameterized post-Newtonian (PPN) parameter $\gamma$ to unprecedented levels of accuracy of $10^{-8}$, it will also reach ability to measure effects of the next post-Newtonian order ($c^{-4}$) of light deflection resulting from gravity's intrinsic non-linearity.  The solar quadrupole moment parameter, $J_2$, will be measured with high precision, as well as a variety of other relativistic effects including Lense-Thirring precession.  LATOR will lead to very robust advances in the tests of fundamental physics: this mission could discover a violation or extension of general relativity, or reveal the presence of an additional long range interaction in the physical law.  There are no analogs to the LATOR experiment; it is unique and is a natural culmination of solar system gravity experiments.

\vspace{1pc}
\end{abstract}

\maketitle

\section{Introduction}

Einstein's general theory of relativity (GR) began with its empirical success in 1915 by explaining the anomalous perihelion precession of Mercury's orbit, using no adjustable theoretical parameters.  Shortly thereafter, Eddington's 1919 observations of star lines-of-sight during a solar eclipse confirmed the doubling of the deflection angles predicted by the theory as compared to Newtonian-like and Equivalence Principle arguments.  From these beginnings, GR has been verified at ever higher accuracy. Thus, microwave ranging to the Viking Lander on Mars yielded accuracy  $\sim$0.1\% in the tests of GR \cite{viking_shapiro2}. The astrometric observations of quasars on the solar background performed with Very-Long Baseline Interferometry (VLBI) improved the accuracy of the tests of gravity to $\sim$ 0.03\% \cite{vlbi_gamma}. Lunar Laser Ranging (LLR)  
	provided $\sim$ 0.01\% verification of GR via precision measurements of the lunar orbit \cite{[39],[62],[63]}.
Finally, the recent experiments with the Cassini spacecraft improved the accuracy of the tests to $\sim$ 0.0023\% \cite{cassini_ber}. As a result, GR has become the standard theory of gravity when astrometry and spacecraft navigation are concerned. 

However, the continued inability to merge gravity with quantum mechanics and recent cosmological data indicate that the pure tensor gravity of GR may need modification or augmentation.  Recent work in scalar-tensor extensions of gravity which are consistent with present cosmological models \cite{[12a],DPV02} motivate new searches for very small deviations of relativistic gravity in the solar system, at levels of 10$^{-5}$ to 10$^{-7}$ of the post-Newtonian effects or essentially to achieve accuracy comparable with the size of the effects of the 2nd order in the gravitational field strength.  This  requires few orders of magnitude improvement in experimental precision from present tests. 

When the light deflection in solar gravity is concerned, the magnitude of the 1st order effect, as predicted by GR for the light ray just grazing the limb of the Sun, is $\sim1.75$ arcsecond. The effect varies inversely with the impact parameter. The 2nd order term is almost six orders of magnitude smaller resulting in  $\sim 3.5$ microarcseconds ($\mu$as) light deflection effect, and which falls off inversely as the square of the light ray's impact parameter \cite{second_order}. The small magnitudes of the effects emphasize the fact that, among the four forces of nature, gravitation is the weakest interaction; it acts at very long distances and controls the large-scale structure of the universe, thus, making the precision tests of gravity a very challenging task. The ability to measure the 1st order light deflection term at the accuracy comparable with the effects of the 2nd order is of the utmost importance for the gravitational theory and is the challenge for the 21st century fundamental physics. 

The LATOR  mission concept will directly address these challenges \cite{space2003}. The test will be performed in the solar gravity field using optical interferometry between two micro-spacecraft.  Precise measurements of the angular position of the spacecraft will be made using a fiber coupled multi-chanelled optical interferometer on the ISS with a 100 m baseline. The primary objective of the LATOR Mission will be to measure the gravitational deflection of light by the solar gravity to accuracy of 0.1 picoradians, which corresponds to $\sim$10 picometers (pm) on a 100 m baseline. 

In conjunction with laser ranging among the spacecraft and the ISS, LATOR will allow measurements of the gravitational deflection by a factor of more than 3,000 better than had recently been accomplished with the Cassini spacecraft. In particular, this mission will not only measure the key PPN parameter $\gamma$ to unprecedented levels of accuracy of 10$^{-8}$, it will also reach ability to measure the next post-Newtonian order ($c^{-4}$) of light deflection resulting from gravity's intrinsic non-linearity. As a result, this experiment will measure values of other PPN parameters such as $\delta$ to $10^{-3}$ (never measured before), the solar quadrupole moment parameter $J_2$ (which is sized at 0.2 $\mu$as using theoretical value of $J_2\simeq10^{-7}$) to 1 part in 20, and the frame dragging effects (the nominal effect is $\pm 0.7 ~\mu$as) on light due to the solar angular momentum to precision of $10^{-2}$.

Technologically LATOR is a very sound concept as all the technologies that are needed for its success have been already demonstrated as a part of the JPL's Space Interferometry Mission (SIM) development.		
	The LATOR concept arose from several developments at NASA and JPL that initially enabled optical astrometry and metrology, and also led to developing expertize needed for the precision gravity experiments. Technology that has become available in the last several years such as low cost microspacecraft, medium power highly efficient solid state and fiber lasers, and the development of long range interferometric techniques make possible an unprecedented factor of 3,000 improvement in this test of GR possible. This mission is unique and is the natural next step in solar system gravity experiments which fully exploits modern technologies.

This paper organized as follows: Section \ref{sec:sci_mot} provides more information about the theoretical framework, the PPN formalism, used to describe the gravitational experiments in the solar system. This section also summarizes the science motivation for the precision tests of gravity that recently became available.  Section \ref{sec:lator_description} provides an overview for the LATOR experiment. Section \ref{sec:conc} discusses the next steps in the mission development.

\section{Scientific Motivation}
\label{sec:sci_mot}
\subsection{PPN Theoretical Framework}

Generalizing on a phenomenological parameterization of the gravitational metric tensor field which Eddington originally developed for a special case, a method called the 
	PPN metric has been developed (see \cite{[39],[40],[59],[44],[58]}).
This method  represents the gravity tensor's potentials for slowly moving bodies and weak interbody gravity, and it is valid for a broad class of metric theories including general relativity as a unique case.  The several parameters in the PPN metric expansion vary from theory to theory, and they are individually associated with various symmetries and invariance properties of underlying theory.  Gravity experiments can be analyzed in terms of the PPN metric, and an ensemble of experiments will determine the unique value for these parameters, and hence the metric field, itself.

In locally Lorentz-invariant theories the expansion of the metric field for a single, slowly-rotating gravitational source in PPN parameters given as
{}
\begin{eqnarray}
g_{00}\hskip -6pt &=&\hskip -5pt 1-2\frac{M}{r}Q(r,\theta) +2\beta\frac{M^2}{r^2}+{\cal O}(c^{-6}),\nonumber\\ 
g_{0i}\hskip -6pt &=&\hskip -5pt  2(\gamma+1)\frac{[\vec{J}\times \vec{r}]_i}{r^3}+
{\cal O}(c^{-5}),\\ 
g_{ij}\hskip -6pt &=&\hskip -9pt -~\delta_{ij}\Big[1+
2\gamma \frac{M}{r} Q(r,\theta)+
\frac{3}{2}\delta \frac{M^2}{r^2}\Big]
\hskip -1pt +\hskip -0pt {\cal O}(c^{-6}),\nonumber
\end{eqnarray}

\noindent where $M$ and $\vec J$ being the mass and angular momentum of the Sun, $Q(r,\theta)=(1-J_2\frac{R^2}{r^2}\frac{3\cos^2\theta-1}{2})$ with $J_2$ being the quadrupole moment of the Sun with $R$ being its radius. $r$ is the distance between the observer and the center of the Sun.  $\beta, \gamma, \delta$ are the PPN parameters and in GR they are all equal to l. The $M/r$ term in the $g_{00}$ equation is the Newtonian limit; the terms multiplied by the post-Newtonian parameters $\beta, \gamma$,  are post-Newtonian terms. The term multiplied by the post-post-Newtonian parameter  $\delta$ also enters the calculation of the relativistic light deflection.

The PPN formalism has proved to be a versatile method to plan gravitational experiments in the solar system and to analyze the data obtained \cite{[39],[40],[59],[44],[58]}. Different experiments test different combinations of PPN parameters (see \cite{[59]}). Using the recent Cassini result \cite{cassini_ber} of $\gamma-1=(2.1\pm2.3)\times 10^{-5}$, the parameter $\beta$ was measured as $\beta-1=(0.9\pm1.1)\times 10^{-4}$ from LLR. The parameter $\delta$ has not yet been measured though its value can be inferred from other measurements.

The technology has advanced to the point that one can consider carrying out tests in a weak field to 2nd order in the field strength parameter. Although any measured anomalies in 1st or 2nd order metric gravity potentials will not determine strong field gravity, they would signal that modifications in the strong field domain will exist.  The converse is perhaps more interesting:  if to high precision no anomalies are found in the lowest order metric potentials, and this is reinforced by finding no anomalies at the next order, then it follows that any anomalies in the strong gravity environment are correspondingly quenched under all but exceptional circumstances. 

We shall discuss the recent motivations for the precision gravity tests below in more details.

\subsection{Motivations for New Gravity Tests}
\label{sec:mot}


The continued inability to merge gravity with quantum mechanics suggests that the pure tensor gravity of GR needs modification or augmentation. The tensor-scalar theories of gravity, where the usual general relativity tensor field coexists with one or several long-range scalar fields, are believed to be the most promising extension of the theoretical foundation of modern gravitational theory. The superstring, many-dimensional Kaluza-Klein, and inflationary cosmology theories have revived interest in the so-called `dilaton fields', i.e. neutral scalar fields whose background values determine the strength of the coupling constants in the effective four-dimensional theory. The importance of such theories is that they provide a possible route to the quantization of gravity and unification of physical law. Although the scalar fields naturally appear in the theory, their inclusion predicts different relativistic corrections to Newtonian motions in gravitating systems. These deviations from GR lead to a violation of the Equivalence Principle (either weak or strong or both), modification of large-scale gravitational phenomena, and generally lead to space and time variation of physical ``constants.'' As a result, this progress has provided new strong motivation for high precision relativistic gravity tests.

The recent theoretical findings suggest that the present agreement between Einstein's theory and experiment might be naturally compatible with the existence of a scalar contribution to gravity. Damour and Nordtvedt \cite{[12a]} (see also \cite{DPV02} for the recent summary of a dilaton-runaway scenario and references therein) have found that a scalar-tensor theory of gravity may contain a `built-in' cosmological attractor mechanism towards GR.  
	Their speculation assumes that the parameter  $\frac{1}{2}(1-\gamma)$  was of order of 1 in the early universe, at the time of inflation, and has evolved to be close to, but not exactly equal to, zero at the present time.
	The numbers for $1-\gamma$ under the various attractor scenarios need checking on under the `new' versions of the mechanism \cite{DPV02} in which the scalar fields' migrations toward the attractor start back at the time of inflation. The expected deviation from zero may be of order of the inverse of the redshift of the time of inflation, or somewhere between $10^{-5}$ and $10^{-7}$ depending on the total mass density of the universe:  $1-\gamma \sim 7.3 \times 10^{-7}(H_0/\Omega_0^3)^{1/2}$, where $\Omega_0$ is the ratio of the current density to the closure density and $H_0$ is the Hubble constant in units of 100 km/sec/Mpc. As a result, this recent work in scalar-tensor extensions of gravity 
	motivate new searches for very small deviations of  gravity in the solar system, at levels of 10$^{-5}$ to $\sim 5\times 10^{-8}$ of the post-Newtonian effects.  


There is now multiple evidence indicating that 70\% of the critical density of the universe is in the form of a ``negative-pressure'' dark energy component; there is no understanding as to its origin and nature. The fact that the expansion of the universe is currently undergoing a period of acceleration now seems well confirmed: it is directly measured from the light-curves of several hundred type Ia supernovae \cite{[2c]}, and independently inferred from observations of CMB \cite{[20],[4c]}. Cosmic speed-up can be accommodated within general relativity by invoking a mysterious cosmic fluid with large negative pressure, dubbed dark energy. The simplest possibility for dark energy is a cosmological constant; unfortunately, the smallest estimates for its value are 55 orders of magnitude too large (for reviews see \cite{[8c]}). Most of the theoretical studies operate in the shadow of the cosmological constant problem, the most embarrassing hierarchy problem in physics. This fact has motivated a host of other possibilities, most of which assume $\Lambda=0$, with the dynamical dark energy being associated with a new scalar field. However, none of these suggestions is compelling and most have serious drawbacks. Given the challenge of this problem, a number of authors \cite{[8c],[17c]} considered the possibility that cosmic acceleration is not due to some kind of new physics, but rather arises from new gravitational physics.

The PPN parameter $\gamma$ may be the key parameter that holds the answers to most of the questions discussed above. In particular, extensions to GR were shown to predict an experimentally consistent universe evolution  without the need for dark energy. These dynamical models are expected to produce measurable contribution to the parameter $\gamma$  in experiments conducted in the solar system also at the level of $1-\gamma \sim 10^{-7}-10^{-9}$, thus further motivating the relativistic gravity research. By testing gravity at ever higher accuracy, one not simply discriminates among the alternative theories of gravity; in effect, one obtains the critical information on the beginning, current evolution and ultimate future of our universe. 

This unexpected discovery demonstrates the importance of testing the new ideas about the nature of gravity. We are presently in the ``discovery'' phase of this new physics, and while there are many theoretical conjectures as to the origin of a non-zero $\Lambda$, it is essential that we exploit every available opportunity to elucidate the physics that is at the root of the observed phenomena. 

In summary, there are a number of theoretical reasons to question the validity of GR. 
	The LATOR mission is designed to provide the needed experimental support to this quest. 

\subsection{PPN Parameters in the Near Future}

Prediction of possible deviation of PPN parameters from the general relativistic values provides a robust theoretical paradigm and constructive guidance for experiments that would push beyond the  empirical upper bound on  $\gamma$ \cite{cassini_ber}  of $\gamma-1=(2.1\pm2.3)\times 10^{-5}$.  
 {}   
Note that the Eddington parameter $\gamma$, whose value in GR is unity, is perhaps the most fundamental PPN parameter, in that $(1-\gamma)$ is a measure, for example, of the fractional strength of the scalar gravity interaction in scalar-tensor theories of gravity.  Within perturbation theory for such theories, all other PPN parameters to all relativistic orders collapse to their GR values in proportion to $(1-\gamma)$. Therefore, measurement of the 1st order light deflection effect at the level of accuracy comparable with the 2nd-order contribution would provide the crucial information separating alternative scalar-tensor theories of gravity from GR discussed in (\cite{[40]}). 

Tests of fundamental gravitational physics feature prominently among NASA's goals, missions, and programs. Among the NASA missions that will study the nature of gravity, we discuss here the missions most relevant to LATOR science:

The Gravity Probe-B (GP-B) mission will be able to measure the geodetic precession effect to $2\times10^{-5}$ and to measure the frame-dragging effect to $3\times10^{-3}$. As a result, this experiment will permit direct measurement of the parameter $\gamma$  with accuracy of $\sim 5\times 10^{-5}$. While the LATOR's accuracy to measure the Lense-Thirring precession will be limited by the small value of the solar angular momentum, the PPN parameter $\gamma$ will be measured with accuracy of almost 3.5 orders of magnitude better than that expected with GP-B.  


LLR contribution to the relativistic tests of gravity comes from its ability to study the lunar orbit to a high accuracy. As such, LLR is basically a $\beta$ experiment rather than a $\gamma$ experiment, primarily testing the non-linearity of gravity theory.
	LATOR will benefit from the technologies developed for LLR over the more than 35 years history of this experiment. However, LATOR will be able to directly measure the PPN parameter $\gamma$ with accuracy of almost 4 orders of magnitude better than currently available. 

	A lander on Mars, with a Cassini-class communication system, may be capable to measure $\gamma$ with accuracy of 10$^{-6}$, limited by the asteroid belt noise.
	As oppose 
ranging to Mars, LATOR  will not be affected by the 
	asteroid modeling problem. Redundancy of the optical truss  and a particular geometry makes LATOR insensitive to such a noise, allowing it to outperform such a mission by at least two orders of magnitude. 

An ambitious test of one of the foundations of GR -- the Equivalence Principle (EP) -- is proposed for the STEP (Space Test of Equivalence Principle) mission.
	STEP will test the composition independence of gravitational acceleration for laboratory-sized bodies
by searching for a violation of the EP with a fractional acceleration accuracy of $\Delta a/a\sim 10^{-18}$ \cite{step2}. 
	The STEP mission will be able to test very precisely for any non-metric, long range interactions in physical law, however the results of this mission will say nothing about the metric component of gravity itself. LATOR is designed specifically to test the metric nature of the gravitational interaction. It will be able to test a number of relativistic effects predicted by the metric gravity and will significantly improve the accuracy for several of these tests. In particular, LATOR will test the key PPN parameter $\gamma$ to unprecedented levels of accuracy of $10^{-8}$, it will also reach ability to measure effects of the order ($c^{-4}$) of light deflection resulting from gravity's intrinsic non-linearity
(further testing the metric structure of the gravity field); it will provide the value for the solar quadrupole moment parameter, $J_2$ 
(very important for the solar-terrestrial physics and gravitational wave experiments in the solar system, but never measured).

A major advantage of the LATOR mission concept is its independence on both -- the drag-free spacecraft environment and the phase-coherent laser transponding techniques. 
	The LATOR experiment is optimized for it's primary science goal -- to measure deflection of light in the solar gravity accurate to 10$^{-8}$. There is no technological breakthroughs needed to satisfy the LATOR mission requirements. All the required technologies exist and most are space-qualified as a part of our on-going interferometry program at JPL.


We shall now discuss LATOR in more details.

\section{Overview of LATOR}
\label{sec:lator_description}

\begin{figure*}[t!]
 \begin{center}
\noindent   \vskip -20pt 
\psfig{figure=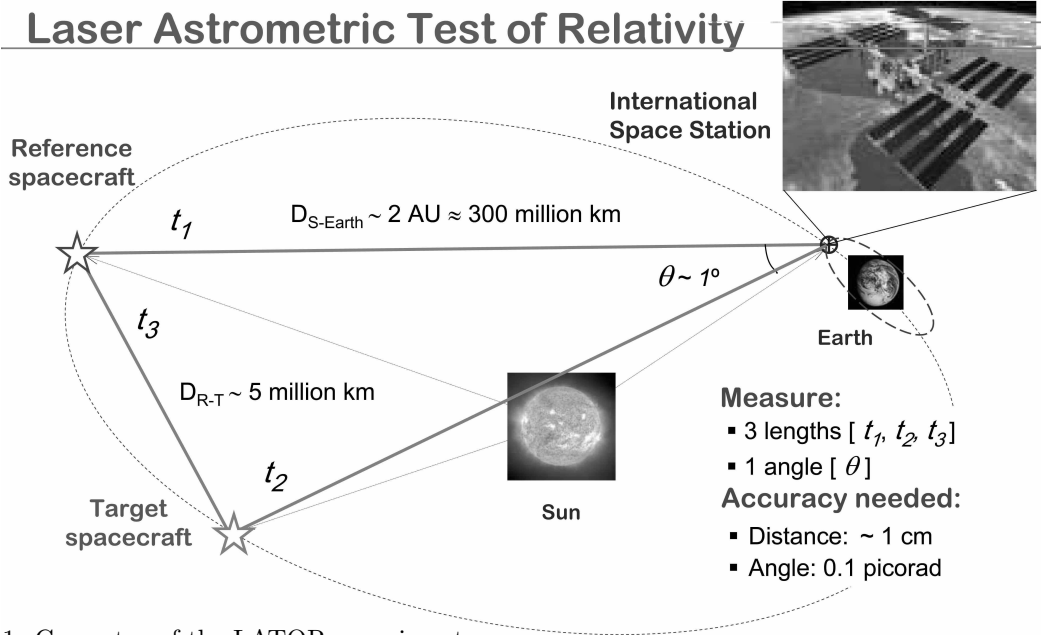,width=138mm}
\end{center}
\vskip -46pt 
  \caption{Geometry of the LATOR experiment.
 \label{fig:lator}}
\vskip -10pt
\end{figure*} 


The LATOR experiment uses laser interferometry between two micro-spacecraft (placed in heliocentric orbits, at distances $\sim$ 1 AU from the Sun) whose lines of sight pass close by the Sun to accurately measure deflection of light in the solar gravity.
	 Another component of the experimental design is a long-baseline ($\sim100$ m) multi-channel stellar optical interferometer placed on the ISS. Figure \ref{fig:lator} shows the general concept for the LATOR mission including the mission-related geometry, experiment details  and required accuracies. 

\subsection{Mission Design}

	The key element of the LATOR experiment is a redundant geometry optical truss to measure the departure from Euclidean geometry caused by Gravity.  The triangle in figure has three independent quantities but three arms are monitored with laser metrology. From three measurements one can calculate the Euclidean value for any angle in this triangle.  In Euclidean geometry these measurements should agree to high accuracy.  This geometric redundancy enables LATOR to measure the departure from Euclidean geometry caused by the solar gravity field to a very high accuracy. The difference in the measured angle and its Euclidean value is the non-Euclidean signal. 
	
	On the ISS, all vibrations can be made common mode for both ends of the interferometer by coupling them by an external laser truss. 
	Additionally, the orbital motion of the ISS provides variability in the interferometer's baseline projection needed to resolve the fringe ambiguity of the stable laser light detection by an interferometer.

The 1st order effect of light deflection in the solar gravity caused by the solar mass monopole is 1.75 arcseconds, which corresponds to a delay of $\sim$0.85 mm on a 100 m baseline. We currently are able to measure with laser interferometry distances with an accuracy (not just precision but accuracy) of $<$ 1 pm. In principle, the 0.85 mm gravitational delay can be measured with $10^{-9}$ accuracy versus $10^{-5}$ available with current techniques. However, we use a conservative estimate for the delay of 10 pm which would produce the measurement of $\gamma$ to accuracy of $10^{-8}$ rather than $10^{-9}$. The 2nd order light deflection is 3.5 $\mu$as (or $\sim$ 1702 pm) and with 10 pm accuracy it could be measured with accuracy of $10^{-3}$, including first ever measurement of the PPN parameter $\delta$.  The frame dragging effect is $\pm0.7~ \mu$as (or $\pm$ 339 pm) would be measured with $10^{-2}$ accuracy and the solar quadrupole moment (using theoretical value $J_2\simeq10^{-7}$) can be modestly measured to 1 part in 20, all with respectable signal to noise ratios.

The laser interferometers use $\sim$2W lasers and $\sim$20 cm optics for transmitting the light between spacecraft. Solid state lasers with single frequency operation are readily available and are relatively inexpensive.   Assume the lasers are ideal monochromatic sources and the measured lengths are 2AU = $3\times 10^8$ km. The beam spread is 1 $\mu$m/20 cm = 5 $\mu$rad. The beam at the receiver is $\sim$1,500 km in diameter, a 20 cm receiver will detect $1.71 \times 10^2$ photons/sec assuming 50\% q.e. detectors. 5 pm resolution for a measurement of $\gamma$ to $10^{-8}$ is possible with $\approx$ 10 sec of integration.

We now outline the basic elements of the LATOR trajectory and optical design.

\subsection{A 3:2 Earth Resonant Orbit}

The two LATOR spacecraft will be launched into the orbit with a 3:2 resonance  with the Earth \cite{teamx}. For this orbit, in 13 months after the launch, the spacecraft are within $\sim10^\circ$ of the Sun with first occultation occuring in 15 months after launch.
	At this point, LATOR is orbiting at a slower speed than the Earth, but as LATOR approaches its perihelion, its motion in the sky begins to reverse and the spacecraft is again occulted by the Sun 18 months after launch.  As the spacecraft slows down and moves out toward aphelion, its motion in the sky reverses again and it is occulted by the Sun for the third and final time 21 months after launch. 

\begin{figure}[t!]
 \begin{center}
\noindent \vskip -8pt    
\psfig{figure=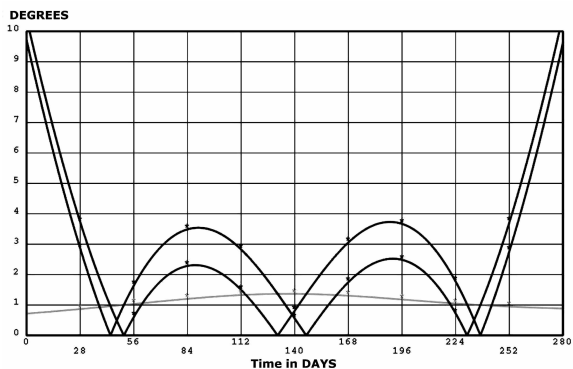,width=76mm}
\end{center}
\vskip -36pt 
  \caption{The Sun-Earth-Probe angle during the period of 3 occultations (two periodic curves) and the angular separation of the spacecraft as seen from the Earth (lower smooth line). Time shown is days from the moment when one of the spacecraft is at 10$^\circ$ distance from the Sun. 
 \label{fig:lator_traj2}}
\vskip -20pt 
\end{figure} 



Figures \ref{fig:lator_traj2} shows the occultations in more details, giving the trajectory when the spacecraft would be within 10$^\circ$ of the Sun as viewed from the Earth.  The two similar periodic curves give the Sun-Earth-Probe angles for the 2 spacecraft while the lower smooth curve gives the angular separation of the spacecraft as seen from the Earth.  

\subsection{Optical Design}

A single aperture of the interferometer on the ISS consists of three 10 cm diameter telescopes. One of the telescopes with a very narrow bandwidth laser line filter in front and with an InGAs camera at its focal plane, sensitive to the 1.3 $\mu$m laser light, serves as the acquisition telescope to locate the spacecraft near the Sun.

The second telescope emits the directing beacon to the spacecraft. Both spacecraft are served out of one telescope by a pair of piezo controlled mirrors placed on the focal plane. The properly collimated laser light ($\sim$10W) is injected into the telescope focal plane and deflected in the right direction by the piezo-actuated mirrors. 

The third telescope is the laser light tracking interferometer input aperture which can track both spacecraft at the same time. To eliminate beam walk on the critical elements of this telescope, two piezo-electric stages are used to move two single-mode fiber tips on a spherical surface while maintaining focus and beam position on the fibers and other optics. Dithering at a few Hz is used to make the alignment to the fibers and the subsequent tracking of the two spacecraft completely automatic. The interferometric tracking telescopes are coupled together by a network of single-mode fibers whose relative length changes are measured internally by a heterodyne metrology system to an accuracy of less than 10 pm.

The spacecraft  are identical in construction and contain a relatively high powered (2 W), stable (2 MHz per hour $\sim$  500 Hz per second), small cavity fiber-amplified laser at 1.3 $\mu$m. Three quarters of the power of this laser is pointed to the Earth through a 20 cm aperture telescope and its phase is tracked by the interferometer. With the available power and the beam divergence, there are enough photons to track the slowly drifting phase of the laser light. The remaining part of the laser power is diverted to another telescope, which points towards the other spacecraft. In addition to the two transmitting telescopes, each spacecraft has two receiving telescopes.  The receiving telescope on the ISS, which points towards the area near the Sun, has laser line filters and a simple knife-edge coronagraph to suppress the Sun light to $10^{-6}$ of the light level of the light received from the space station. The receiving telescope that points to the other spacecraft is free of the Sun light filter and the coronagraph.

In addition to the four telescopes they carry, the spacecraft also carry a tiny (2.5 cm) telescope with a CCD camera. This telescope is used to initially point the spacecraft directly towards the Sun so that their signal may be seen at the space station. One more of these small telescopes may also be installed at right angles to the first one to determine the spacecraft attitude using known, bright stars. The receiving telescope looking towards the other spacecraft may be used for this purpose part of the time, reducing hardware complexity. Star trackers with this construction have been demonstrated many years ago and they are readily available. A small RF transponder with an omni-directional antenna is also included in the instrument package to track the spacecraft while they are on their way to assume the orbital position needed for the experiment. 

\section{Conclusions}
\label{sec:conc}


The LATOR experiment has a number of advantages over techniques that use radio waves to measure gravitational light deflection. The optical technologies allows low bandwidth telecommunications with the LATOR spacecraft. The use of the monochromatic light enables the observation of the spacecraft almost at the limb of the Sun. The use of narrowband filters, coronagraph optics and heterodyne detection will suppress background light to a level where the solar background is no longer the dominant noise source. The short wavelength allows much more efficient links with smaller apertures, thereby eliminating the need for a deployable antenna. Finally, the use of the ISS enables the test above the Earth's atmosphere -- the major source of astrometric noise for any ground based interferometer. This fact justifies LATOR as a space mission.

The LATOR experiment technologically is a very sound concept; all technologies that are needed for its success have been already demonstrated as a part of the JPL's interferometry program.  
	The LATOR experiment does not need a drag-free system, but uses a geometric redundant optical truss to achieve a very precise determination of the interplanetary distances between the two micro-spacecraft and a beacon station on the ISS. The interest of the approach is to take advantage of the existing space-qualified optical technologies leading to an outstanding performance in a reasonable mission development time.  The  availability of the ISS makes this mission concept realizable in the very near future; the current mission concept calls for a launch as early as in 2009 at a cost of a NASA MIDEX mission.   

LATOR will lead to very robust advances in the tests of fundamental physics: it could discover a violation or extension of GR, or reveal the presence of an additional long range interaction in the physical law.  There are no analogs to the LATOR experiment; it is unique and is a natural culmination of solar system gravity experiments.

The work described here was carried out at the Jet Propulsion Laboratory, California Institute of Technology, under a contract with the National Aeronautics and Space Administration.


\end{document}